\magnification 1200
\rightline{DAMTP/97/56}
\rightline{SPhT/97/56}
\vskip 1 cm
\centerline{\bf EXPONENTIALLY SMALL COUPLINGS BETWEEN }
\vskip .5 cm
\centerline{\bf TWISTED FIELDS  OF ORBIFOLD STRING THEORIES}
\vskip 1 cm
\centerline{by}
\vskip .5 cm
\centerline{{Ph. Brax}\footnote{$^1$}{Present address: SPhT Saclay F-91191 Gif/Yvette cedex France}\footnote{$^{2}$}{brax@spht.saclay.cea.fr}}
\vskip .5 cm
\centerline{and}
\vskip .5 cm 
\centerline{{Neil Turok}\footnote{$^3$}{n.g.turok@amtp.cam.ac.uk}}
\vskip 1 cm
\centerline{Department of Applied Mathematics and Theoretical Physics}
\vskip .2 cm
\centerline{Silver Street}
\vskip .2 cm
\centerline{Cambridge CB39EW}
\vskip .2 cm
\centerline{England}
\vskip 2 cm
\leftline{\bf Abstract:}
\vskip .2 cm
We investigate the natural occurrence of exponentially small couplings in effective field theories deduced from higher dimensional models.
 We calculate the coupling between twisted fields of the $Z_3$ Abelian orbifold compactification of the heterotic string.  Due to the propagation of massive Kaluza-Klein modes between the fixed points of the orbifold, the massless twisted fields located at these singular points become weakly coupled. The resulting small couplings   have an exponential dependence   on the mass of the intermediate states and the distance between the fixed points.

\vfill\eject
\input epsf
\leftline{\bf I Introduction}
\vskip 1 cm
Many phenomena explained by effective field theories require small couplings or small mass ratios.  This is for instance the case of the hierarchy problem where there is a large discrepancy  between the weak scale $m_W$ and the Planck mass. A possible solution to this puzzle would be to introduce very small couplings in the  Lagrangian of the standard model. However these small couplings are very difficult to motivate from the point of view of field theory. Indeed, most low energy effective field theories are expected to be natural, i.e all couplings which are not forbidden for symmetry reasons are of order $O(1)$$^{[1]}$. One would therefore like to investigate the mechanisms which could help  generating very small couplings. A possible scenario can be envisaged when the low energy effective field theories are approximations of more fundamental theories with extra dimensions$^{[2]}$.
The presence of these extra dimensions entails the existence of massive Kaluza-Klein modes upon compactification. These modes  couple fields living far apart in the extra dimensions. These fields become very weakly coupled due to the small correlation length implied by the massive Kaluza-Klein modes. 
 These small couplings appear when the compactifying space exhibits curvature singularities where fields are located.  A natural realisation of this mechanism is provided by the compactification of the heterotic string on an orbifold$^{[3]}$. Indeed the action of the point group of the orbifold on the six dimensional torus has fixed points leading to singularities . Each of these fixed points is associated to massless twisted states. Massive untwisted solitonic strings can propagate between the fixed points$^{[4]}$. This entails the existence of exponentially small couplings between twisted states at different fixed points$^{[5,6]}$. 

 In section II, we consider  the case of a free field theoretic  model compactified on a singular $D$ dimensional torus. The effective interaction between fields situated at two different singularities is exponentially small due to the exchange of one massive mode between the fixed points. In section III we extend our analysis to orbifold string models. Choosing the case of the $Z_3$ orbifold, we show that exponentially small couplings between twisted states are present. We have included an appendix where we derive some useful results on free propagators.   
 \vskip 1 cm
\leftline{\bf II Exponentially Small Couplings and Toroidal Compactification}
\vskip 1 cm 
Let us consider the interaction between a complex scalar field $\Psi$ in
($d+D$) dimensions and two complex scalar fields $\phi_1$ and $\phi_2$
assigned to  two fixed points in the extra $D$ dimensions, i.e. $\phi_{1,2}(x,y)=\phi_{1,2}(x)$ if $y=f_{1,2}$ and zero otherwise where $y$ parametrises the extra dimensions. The  two fixed points $f_{1,2}$ are singularities in the extra dimensions. We assume that the extra dimensions are flat apart from a  curvature singularity at the fixed points, i.e. $g^{1\over 2}R=\delta (x_{d+1}-f_1) +\delta (x_{d+1}- f_2)$. The action involves the kinetic terms
$$S_{kin}= \int d^d x\ \vert\partial_{\mu}\phi_1(x)\vert^2+
\int d^d x\ \vert\partial_{\mu}\phi_2(x)\vert^2 +\int d^{d+D}x
\Psi^*(-\partial_A^2-M^2)\Psi\eqno(1)$$
where $A=1..(d+D)$ and the Minkowsky space-time index $\mu=1..d$. The
mass $M$ is the mass of the scalar field $\Psi$. The interation term
reads
$$S_{int}=-\lambda\Lambda^{{(4+D-d)}\over 2}\int d^{d+D} x \sqrt g R\Psi(\phi_1^2 
+ \phi_2^2)+hc\eqno(2)$$
where $\Lambda$ is the scale where the model is defined and $\lambda$ a positive dimensionless constant. We  assume that the  extra dimensions are curled up on a torus.
 This amounts to compactifying the model on a lattice $2\pi L$ where $L$ is integral. This can serve as a prelude to the string calculation.
The scattering amplitude $\phi_1+\phi_1\to \phi_2+\phi_2$ is given by the propagator of the massive particle propagating between the two singularities $T=-{\lambda^2\Lambda^{4+d-d}\over 2}G$.
We
shall derive several useful representations for the propagator
$$G=<f_2\vert ({{M^2-s-\Delta_{D}}\over 2})^{-1}\vert
f_1>\eqno(3)$$ 
which will reappear in an appropriate guise in string theory.

First of all, let us use the completeness relation $\sum_{p\in L^*}
\vert p><p\vert =1$ where the sum is over the dual lattice $L^*=\{p/ \
p.x\in  Z, z\in L\}$. This yields
$$<f_2\vert ({{M^2-s-\Delta_D^2}\over 2})^{-1}\vert
f_1>={V_L^{-1} }\int_0^{\infty}dt_E e^{-{{(M^2-s)}\over
2}t_E} \sum_{p\in L^*}e^{2 \pi ip.(f_1-f_2)}e^{-{{p^2}\over 2}t_E}\eqno(4)$$
where $\vert p>$ is an eigenstate of $-\Delta_D$
and
$$<f_2\vert e^{t_E{{\Delta}\over 2}}\vert f_1>= 
V_L^{-1}\sum_{p\in L^*}e^{2\pi ip.(f_1-f_2)}e^{-{{p^2}\over 2}t_E}\eqno(5)$$.
The propagator corresponds to the propagation of a
Gaussian wave
packet in Euclidean time from $f_1$ to $f_2$. Let us define 
the Hilbert space $\cal H$ spanned by the eigenstates $\vert p>$. 
One can reformulate the heat kernel as
$$<f_2\vert e^{t_E{{\Delta_D}\over 2}}\vert f_1>=Tr_{\cal
H}(e^{2\pi iP.(f_1-f_2)}e^{-t_E H})\eqno(6)$$
where $P=i\nabla$ is the momentum operator and $H={P^2\over 2}$ is the
Hamiltonian. This Hamiltonian version of the propagator can be 
 written as a path integral over classical trajectories.
Indeed,
we can use the Poisson resummation formula to obtain a sum over the
lattice $L$. Using 
$$e^{-{{p^2t_E}\over 2}}={1\over (2\pi t_E )^{D\over 2}}\int d^D
xe^{ip.x}e^{-{x^2 \over 2t_E}}\eqno(7)$$
and the resummation formula
$$\sum_{p\in L^*}e^{ip.x}={V_L}\sum_{q\in L}\delta (x-2\pi
q)\eqno(8)$$
one gets 
$$G=
\sum_{q\in L}\int_0^{\infty} dt_E{1\over (2\pi t_E)^{D\over 2}} 
e^{-t_E{{(M^2-s)}\over 2}}e^{-(2\pi)^2{{(f_1-f_2-q)}\over 2t_E}}
\eqno(9)$$
Finally, this is equivalent to the path integral 
$$G=
 \sum_{q\in L}\int_0^{\infty}dt_Ee^{-t_E{{(M^2-s)}\over 2}} 
 \int_{2\pi f_1}^{2\pi f_2+2\pi q} 
{\cal D}xe^{-{1\over 2}\int_0^{t_E}\dot x^2d\tau}\eqno(10)$$
The sum is due to the periodicity of the lattice. 
If $s<M^2$ this integral converges.
In
the vicinity of $s=M^2$, there is a pole and the propagator is
dominated by the $p=0$ term in (4)
$$<f_2\vert{1\over M^2-s-\Delta_D}\vert f_1>\sim {V_L^{-1}\over
M^2-s}\eqno(11)$$
 This
comes from the fact that the propagator is solution of
$$(M^2-s-\Delta_D)G=\sum_{q\in L}\delta (x-2\pi q)\eqno(12)$$
Using the resummation formula (8) and writing $G(x)=\sum_{p\in
L^*}G_pe^{ip.x}$ one gets
$$\sum_{p\in L^*}(m^2-s+p^2)G_pe^{ip.x}=V_L^{-1}\sum_{p\in L^*}e^{ip.x}\eqno(13)$$
It is easy to see that $G_0={V_L^{-1}\over M^2-s}$. Notice that
the pole disappears as $V_L\to\infty$, it is a finite size effect. 
Below $M^2$, the propagator receives contributions from the
whole series. In fact, one can use a saddle point
approximation in this regime (more details are given in the appendix)
As a result,
 the propagator reads
$$<f_2\vert{1\over m^2-s-\Delta_D}\vert f_1>\sim {1\over 
2^D(M^2-s)^{D\over 2}}\sum_{q\in L}e^{-2\pi 
\sqrt {M^2-s}\vert f_2-f_1+q\vert}\eqno(14)$$
below the pole. 
This result specialised for $s\ll M^2$ gives the value   of the coupling between the two fixed points
$$V_{coup}=-{{\lambda^2\Lambda^{4+D-d}}\over (2M)^D}\sum_{q\in L}e^{-2\pi M\vert f_2-f_1+q\vert}(\phi_1^2\phi_2^{*2}+c.c.)\eqno(15)$$
 One can immediately generalise this result to the gauge model where the field $\phi_{1,2}$ transform in the vectorial representation of a gauge group G
$$V_{coup}=-{{\lambda^2\Lambda^{4+D-d}}\over (2M)^D}\sum_{q\in L}e^{-2\pi M\vert f_2-f_1+q\vert}(\bar\phi_1 T^a\phi_1)(\bar \phi_2 T^a \phi_2)\eqno(16)$$
where $T^a$ are the generators of the gauge group. 
We see that the compactification on a singular torus leads to exponentially small couplings in the effective theory describing the uncompactified dimensions.
\vskip 1 cm
\leftline{\bf II Small Couplings in Orbifold String Theories}
\vskip 1 cm
Orbifold string theories provide an elegant and tractable scheme where stringy calculations can be performed$^{[7]}$. Moreover, there are singularities on an orbifold which are of the type discussed in the previous section, i.e. four dimensional twisted states are assigned to the singularities of the  orbifold.
It  seems natural to look for small couplings between states living far apart at different fixed points. The analogue of the massive field $\Psi$ is the tower of untwisted states which can propagate between fixed points.
\vskip 1 cm
\midinsert
\epsfysize=8 cm
{\centerline{\epsfbox{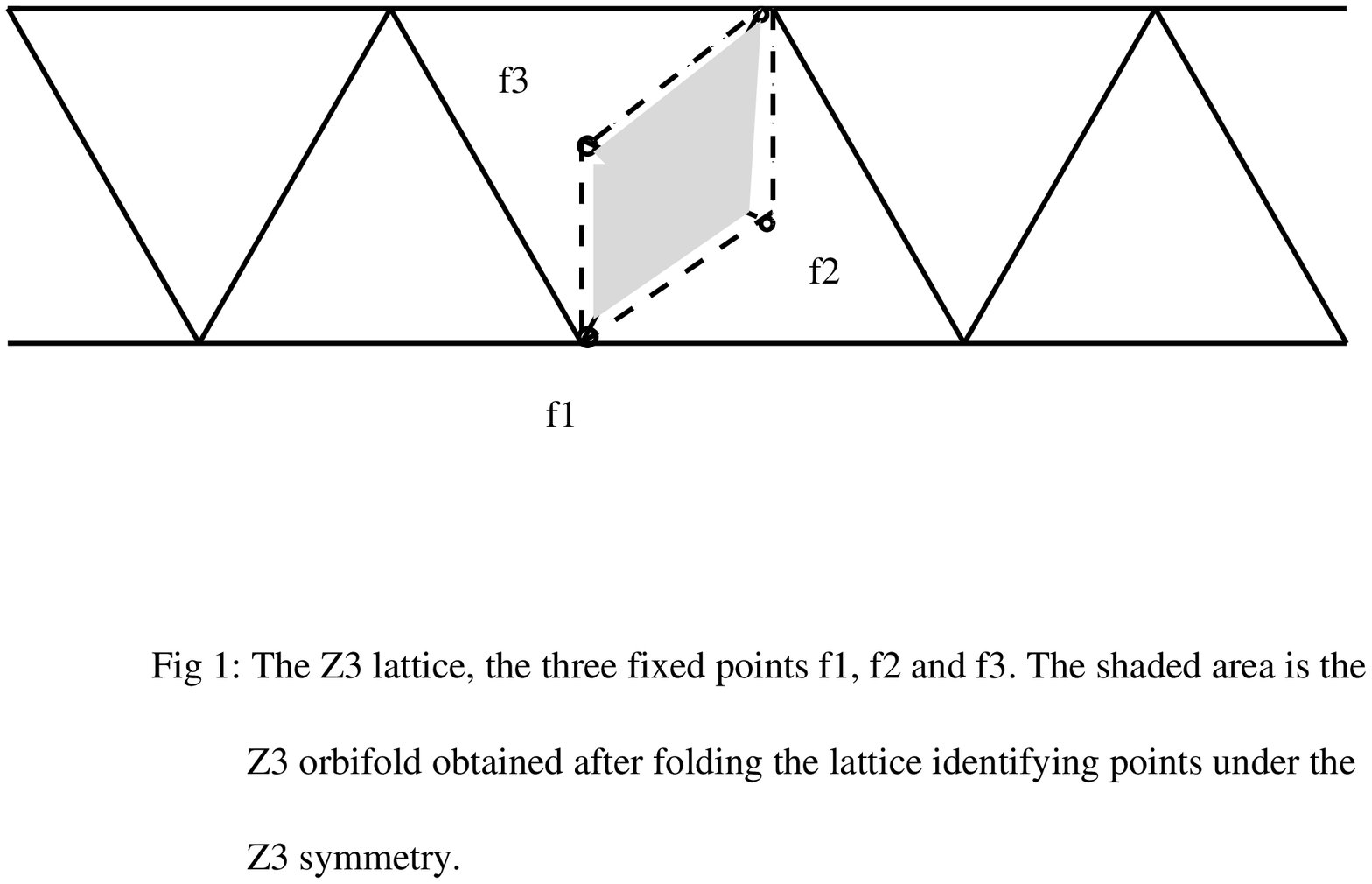}}}
\endinsert
\vskip .5 cm
We shall consider the $E_8\times E_8$ heterotic string theory compactified on an Abelian orbifold. The left part of the theory contains the gauge degrees of freedom while the right part is supersymmetric on the world sheet and is responsible for the spin content of the spectrum. The compactification from $10$ to $4$ dimensions is obtained by  considering the theory on a six dimensional torus on which acts an Abelian group of automorphisms of the corresponding lattice $L$. As the automorphism group does not act freely on the torus, the resulting quotient space is not a manifold. There are singularities associated to the fixed points of the group of automorphism. When analysing the massless spectrum of the theory, one finds that two types of states are present. The untwisted states correspond the dimensional reduction of $10$ dimensional states whereas the twisted states correspond to closed strings 
satisfying
$$X_i(\sigma+2\pi)=\theta^jX_(\sigma)+\Lambda_i\eqno(17)$$
where $X_i$ are the six bosonic coordinates on the orbifold, $\theta$ is the generator of the abelian group in the $Z_N$ case and $\Lambda$ is an element of the lattice $L$. These states are in the $j$th twisted sector. Each of these states is associated to one of the fixed points of the orbifold. This implies that the center of mass of these states is at one of the fixed points $f$. Twisted states give a natural realisation of the fields $\phi_{1,2}$ of the previous analysis.

The gauge group of the $(N=1)$ model is determined by the left part of the orbifold string theory. The left part contains $16$ scalar fields or equivalently $16$ complex 
fermions which can be bosonised on the $E_8\times E_8$ lattice
$\lambda_i\sim e^{i\beta_I^i F_I}$ where $i=1..16$, $I=1..16$ and $\beta=(\underline {\pm 1, 0,0....,0,0})$. The gauge bosons are determined by $e^{iP.F}$ where $P$ is  a root of the gauge lattice, i.e. $P^2=2$. The generator of the abelian group automorphisms $\theta$ acts simultaneously on the gauge lattice by translation. The left part of the gauge boson vertex  operator is invariant under the action of $\theta$. This implies that $P.V\in Z$.
The choice of the gauge group can be even more constrained by considering Wilson lines$^{[8]}$ $A^I_i$, $I=1..16,i=1..6$. Integrating the Wilson lines along one of the six cycles of the torus defined by $L$ gives $\int_i dx_m A^m_I=2\pi a^i_I$. The gauge group is  restricted as  the root vectors must satisfy $P.a^i\in Z$.
Wilson lines also modify the twisted spectrum . For the Abelian $Z_3$ orbifold, the lattice vectors determining the spectrum satisfy $(P+nV+ma)^2={4\over 3}$ where $n=0..2$ labels the twisted sector and $m=0..2$ distinguishes the different fixed points. 
We shall use  the $Z_3$  orbifold and its $27$ fixed points. 

We are interested in exponentially small terms in the scattering of four twisted states. Indeed the scattering amplitude between these states is the stringy analogue of the previous field theoretic calculations where fields lived at singular points in the compactified dimensions.  Fixed points being separated by a distance proportional to the orbifold radius, one can expect that the scattering amplitude is exponentially suppressed due to the exchange of massive string modes.
The calculation of the scattering amplitude involving four twisted states gives information on renormalisable interaction terms in four dimensions. Higher order scattering amplitude would lead to  non-renormalisable interaction terms.
The calculation will be done at tree level on the sphere. 
We closely follow the original analysis of Dixon et al. in ref.[5].

 Let us explicitly calculate  the scattering amplitude in the case of a standard embedding of the space group in the gauge lattice $V=({1\over 3},{1\over 3},-{2\over 3},0,0,0,0,0)$.
One can  choose one  twisted field in the
$27 $ of $E_6$ and one
anti-twisted field in the $\bar {27}$ 
living  at the fixed point $f_1$, while two other
$27$ and $\bar {27} $ are at another fixed point $f_2$.
The correlation function $<27(f_2)\bar{27}(f_2)27(f_1)\bar{27}(f_1)>$ 
is not forbidden by the point group selection
rule (two twisted and two anti-twisted fields are $\theta$ invariant) and the space group
selection rule (the combination of fixed points satisfies $f_1+f_2-f_1
-f_2=0$). Using conformal invariance, one can send three points to $0,1$
and $\infty$. The fourth argument of the correlation function is the
cross ratio $x$.  
Let us  compute
$$Z(x)=\mathrel{\mathop {\lim}_{z_{\infty}\to\infty}}\vert
z_{\infty}\vert^4
<\bar{27}_{f_2}(z_{\infty})27_{f_2}(1)27_{f_1}(0)\bar{27}_{f_1}(x)>\eqno(18)$$
where $27_{f_1}(0)$ is the vertex operator of the field in the $27$ of
$E_6$ located at the origin of the sphere and embedded in the
orbifold at the fixed point $f_1$.
Vertex operators are a product of a left part which  contains the gauge degrees of freedom  with a right part involving
the $so(10)$ supersymmetric content of the fields. The vertex operators of a $27$ twisted field looks like
$$27_{L,f_1}(0)= e^{i(P+V).F}\Lambda_{+,f_1}(0)\eqno(19)$$
(respectively $-P-V$ for an anti-twisted $\bar{27}$). The twist operator
is denoted by $\Lambda_{+,f_1}(0)$. The massless states  satisfy
$(P+V)^2={4\over 3}$, i.e. $P$ can be equal to
$$\eqalign{
(0,0,1,\underline{\pm 1,0,0,0,0})\cr
(-{1\over 2},-{1\over 2},{1\over 2},\underline{\pm {1\over
2},0,0,0,0})\cr
(-1,-1,0,0,0,0,0,0)\cr}
\eqno(20)$$
The first possibility builds up the $10$ of $so(10)$ while the second
gives the $16$ and the last is a singlet. 
The left part of the correlation function
(18) reads
$$<e^{i(P_1+V).F(0)}e^{-i(P_2+V).F(\bar x)}e^{-i(P_3+V).F(\bar
z_{\infty})}e^{i(P_4+V).F(1)}>Z_{twist}(\bar z_{\infty})\eqno(21)$$
where 
$$Z_{twist}(\bar z_{\infty})=<\Lambda_{+,f_1}(0)\Lambda_{-,f_1}(\bar x)
\Lambda_{-,f_2}(\bar z_{\infty})\Lambda_{+,f_2}(1)>\eqno(22)$$
The correlation function  is non-zero if $P_1+P_4=P_2+P_3$. 
An easy calculation gives
$$\bar z_{\infty}^{a}\bar x^{b} (1-\bar x)^c Z_{twist}(\bar
z_{\infty})\eqno (23)$$
where
$$\eqalign{
a&=-(P_3+V)^2\cr
b&=-(P_2+V).(P_1+V)\cr
c&=-(P_2+V).(P_4+V)\cr}
\eqno(24)$$
The exponents depend on the choice of combinations of $P$'s.
We shall choose them in such a way that $P_1 -P_2$ 
correspond to the adjoint representation in the product
$$\underline {27}\times
\underline{\overline{27}}=\underline{1}+\underline{78}+\underline{
650}\eqno(25)$$
in the $s$ channel.
The adjoint representation $\underline{78}$ is defined by
the root lattice of $E_6$ such that $(P_1-P_2)^2=2$, for instance 
$P_1=(0,0,1,1,0,0,0,0,0)$ and $P_2=(0,0,1,0,1,0,0,0)$. The exponents
become
$$\eqalign{
a=-{4\over 3}\cr
b=-{1\over 3}\cr
c=-{1\over 3}\cr }
\eqno(26)$$
These exponents are determined  once two constraints are imposed. First of all the limit $x\to 0$ must show a pole in the $s$ channel corresponding to the exchange of a gauge boson. Then the limit $x\to \infty$ must correspond to a non-vanishing Yukawa coupling between three twisted states. For other embeddings  of the space group of the orbifold in the gauge lattice and if Wilson lines are included, the    twisted states belong to various representations of a gauge group G. If one assumes that there exists a gauge invariant combination of three twisted states involving $f_1, f_2$ and another fixed point whose coupling is not forbidden by the space group rule, then the left part of the correlation function is still given by (23).

Let us now deal with the right part of the correlation function.
The vertex operators depend on the picture chosen. In the $-1$ picture
the vertex operators read
$$27_{+,f_1,R,-1}(0)=e^{-\phi(0)}e^{i(\alpha_v+v).H(0)}\Lambda_{+,f_1}(0)
\eqno(27)$$
where $\alpha_v=(0,0,1,0,0)$.
In the $0$ picture, the part of the vertex operator yielding a
non-zero 4-point function depends on the four-momentum of the state
$$\bar
{27}_{-,f_1,R,0}(x)=ik_2.\psi(x)e^{-i(\alpha_v+v).H(x)}\Lambda_{-,f_1}(x)
\eqno(28)$$
We can assign the states at the origin and $1$ to be in the $-1$ picture whereas states at $x$ and $\infty$ are in the $0$ picture  (the sum of the charges is $-2$).
Then using $<\psi_{\mu}(z_{\infty} )\psi_{\nu}(x)>\sim{\eta_{\mu\nu}\over z_{\infty}}$
and
$<e^{-\phi(1)}e^{-\phi(0)}>=1$ as
$\hbox{dim}(e^{-\phi})={1\over 2}$, one gets 
$$k_2.k_3 z_{\infty}^{-{4\over 3}}x^{-{1\over 3}}(1-x)^{-{1\over
3}}Z_{twist}(z_{\infty}) \eqno(29)$$
where we have used $(\alpha_v +v)^2={1\over 3}$ and denoted by
$Z_{twist}(z_{\infty}) $ the right part of the twist correlation
function.
The correlation function reads
$$Z(x)=k_2.k_3\vert 1-x \vert ^{-{2\over 3}-u} \vert x\vert ^{-{2\over 3}-s}Z_{twist} \eqno(30)$$
where
$$Z_{twist}=\mathrel{\mathop {\lim}_{z_{\infty}\to \infty}}\vert 
z_{\infty}\vert^{4\over
3}<\Lambda_{+,f_1}(0)\Lambda_{-,f_1}(x,\bar x)\Lambda_{+,f_2}(z_{\infty},\bar
z_{\infty})\Lambda_{+,f_2}(1)>\eqno(31)$$
and $s=(k_1+k_2)^2$, $u=(k_1-k_4)^2$ are the Mandelstam variables.

The twist correlation function is obtained by calculating the path
integral of bosonic strings on the orbifold satisfying the boundary
conditions prescribed by the four twisted states at $0,1,x$ and
$\infty$. One can separate the bosonic string into a quantum and
a classical part. The classical strings represent untwisted states
propagating between the fixed points. The quantum part corresponds to
the fluctuations around the classical strings, in particular the
quantum part of the correlation function reproduces the leading
exponents in the operator product expansion as $x\to 0,1$, i.e
$Z(x)=Z_{qu}Z_{cl}$ where
$$Z_{qu}=\sin{\pi\over 3}V_L
{{\vert x \vert^{-{4\over 3}}\vert 1-x \vert ^{-{4\over
3}}}\over I^3}\eqno(32)$$ 
More precisely, the quantum part depends on  
the hypergeometric function
$$F(x)={{\sin {\pi\over 3}}\over \pi} \int_{0}^1 dy{1\over
y^{1\over 3}(1-y)^{2\over 3}(1-xy)^{1\over 3}}\eqno(33)$$ 
as $I=F(x)\bar F (1-\bar x)+\bar F (\bar x) F(1-x)$. One also needs to
define  
$\tau=i{F(x)\over F(1-x)}$.
This gives
$$Z(x)=\sin {\pi\over 3}V_L 
k_2.k_3{{\vert 1-x\vert^{-2-u}\vert x\vert^{-2-s}}\over
I^3}Z_{cl}\eqno(34)$$
We are therefore left with the evaluation of the classical action of
untwisted string on the orbifold
$$Z_{cl}=\sum e^{-S}\eqno(35)$$
where the sum is over the set of classical solutions with classical
action $S$. Let us now determine this classical solutions.

 Recall that  the
bosonic string action in the conformal gauge reads
$$S={1\over 4\pi}\int d^2z (\partial X\bar \partial \bar X+\bar
\partial X \partial \bar X)\eqno(36)$$
where $X=(X^1,X^2,X^3)$ is a complex scalar field build from the six
bosonic fields $X_i, i=1..6$ of the orbifold, e.g. $X^1=X_1+iX_2$.
 The variable
$z$ lives on the Riemann sphere.
The path integral yielding the scattering amplitude is dominated by
classical configurations 
 characterised by the Laplace equation
$$\Delta X=0\eqno (37)$$
These instantons are constrained by appropriate boundary conditions.
They are classified according to their degree. Indeed,
they represent maps from the Riemann sphere to the orbifold. As the
orbifold $Z_3$ is homeomorphic to a 2-sphere, one therefore deduces
that the instantons are maps belonging to $\Pi_2(S^2)=Z$. Each of
these instantons is specified by one integer, i.e. its degree. The
degree describes the number of time the closed strings wraps around the
orbifold. We shall see that instantons of any degree exist. 
\vskip 1 cm
\midinsert
\epsfysize=5.cm
\epsfxsize=8 cm
{\centerline{\epsfbox{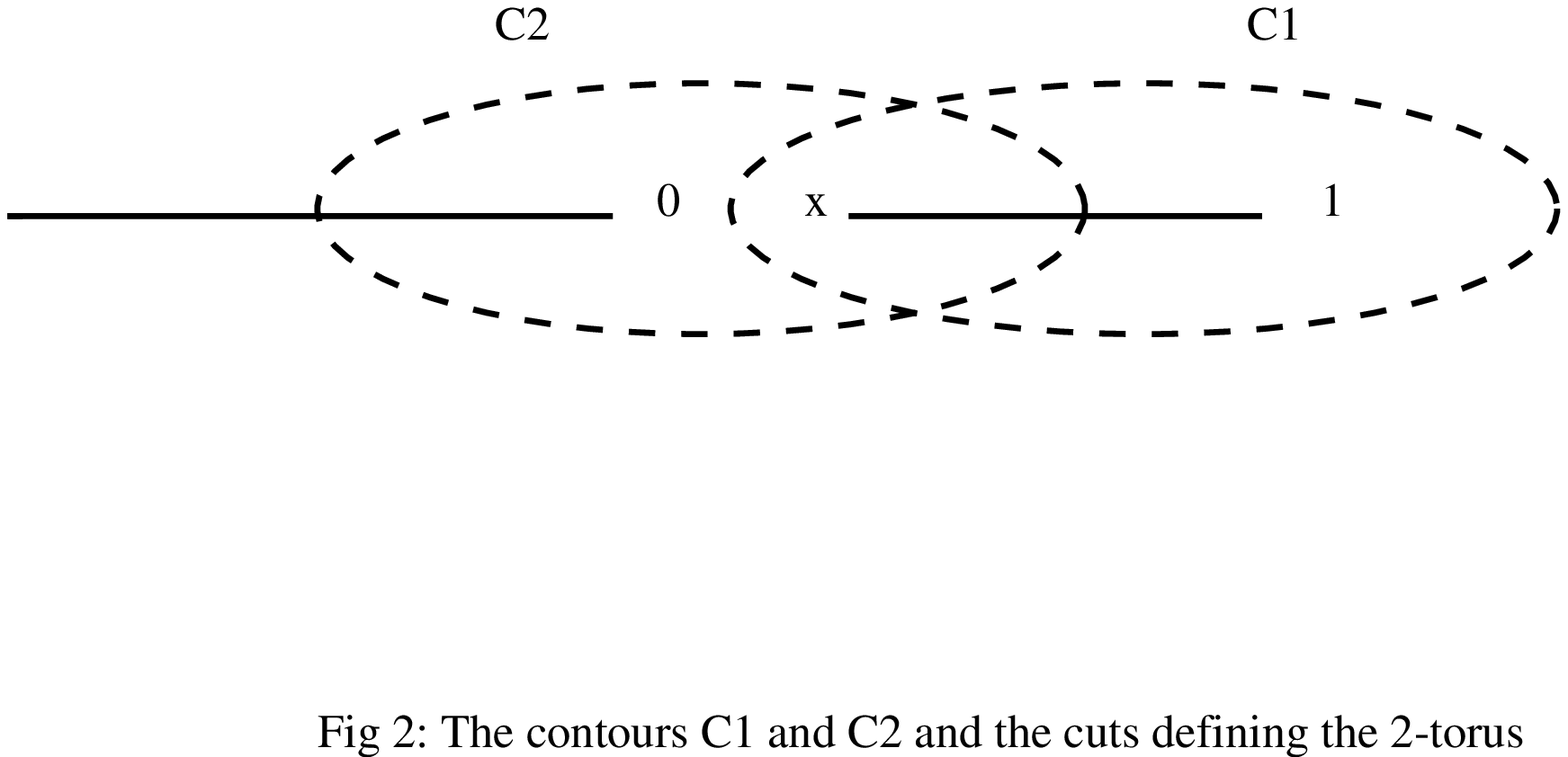}}}
\endinsert
\vskip .5 cm
As the Riemann sphere has four marked points where the twisted states are
attached $0,1,\infty$ and $x$, the instantons are multivalued
functions on the sphere. One can construct a 3-sheeted covering of the
sphere with four branch points of order $3$ where the instanton is single-valued (locally, the twist
field introduce a cubic root close to each branch point). The
corresponding Riemann surface has a genus given by the Riemann-Hurwitz
formula $2g-2=3*(-2) + 4*2$, so the genus is $g=2$.
The solutions of the Laplace equation are obtained once a basis for
the Abelian differentials are known on the 2-torus. The dimension of
the space of Abelian differentials is 2 spanned by
$$\eqalign{
\omega_1={{dy}\over y^{1\over 3}(1-y)^{1\over 3}(x-y)^{2\over 3}}\cr
\omega_2={{dy}\over y^{2\over 3}(1-y)^{2\over 3}(x-y)^{1\over 3}}\cr}
\eqno(38)$$
 The
2-torus is obtained by considering the complex plane with two cuts,
one between $-\infty$ and $0$ along the real axis, the other one
between  $x$ and $1$ along the real axis. 
There are four independent loops on the
two-torus. In fact they  can be all generated from two loops in the
complex plane ${\cal C}_1$ and ${\cal C}_2$ respectively encircling
the cut from $x$ to $1$ and going across the cuts  to encircle the
segment between $0$ and $x$. Four other loops can be obtained by
transferring them to the two other sheets. One can select four of
these loops to form a homology basis  of the 2-torus.

We can now discuss the boundary conditions for the untwisted strings
propagating between the fixed points. These boundary conditions
completely specify the instantons. 
The instantons are determined by their monodromy properties around the
closed loops of the two-torus with net twist zero (they are
untwisted).
Indeed, consider the two twisted strings at $x$ and the origin, they
satisfy
$$\eqalign{
X_0(\sigma+2\pi)=\theta X_0(\sigma)+(1-\theta)(f_1+q_0)\cr
X_x(\sigma+2\pi)=\theta^{-1} X_x(\sigma)+(1-\theta^{-1})(f_1+q_x)\cr}
\eqno(39)$$
when transported along a closed loop surrounding $0$ (respectively
$x$).
When these two twisted string merge to create a single untwisted
closed string, the monodromy of the resulting closed string $X$ when
transported around a closed loop surrounding $x$ and $0$ is obtained
by combining $X(\sigma+2\pi)=\theta (\theta^{-1}X(\sigma)
+(1-\theta^{-1})(f_1+q_x) )+(1-\theta)(f_1+q_0)$. 
The resulting string is untwisted with a winding number
$v_2=(1-\theta)(q_0-q_x)$ (this phenomenon is sketched on fig. (4)). Similarly, one has to deal with boundary
conditions involving the other external strings at $1$ and $\infty$.
As the homology of the 2-torus is spanned by two cycles ${\cal C}_1$
and ${\cal C}_2$, one only needs to study the monodromy of the
untwisted string wrapped along these two contours. The contour ${\cal
C}_2$ corresponds to the previous case. One can  deal with
the other case to obtain 
$$\Delta_{{\cal C}_i} X=\int _{{\cal C}_i}\partial X +\int_{{\cal
C}_i}\bar \partial X= 2\pi v_i\eqno(40)$$
where ${\cal C}_i$ are the independent loops of the 2-torus.
The vectors $v_i$ belong to certain cosets of the lattice $L$ of the
orbifold, $v_1\in (1-\theta)(f_2-f_1+L),\ v_2\in (1-\theta)L$.

The instantons are solutions of
$$\eqalign{
\partial X=a\omega_1\cr
\bar \partial X=b\bar\omega_2\cr}
\eqno(41)$$
Notice that $\omega_1$ and $\bar \omega_2$ pick up a
$\theta=e^{i{{2\pi}\over 3}}$ phase factor when crossing the cuts. We
only need to check the monodromies for the loops ${\cal C}_1$ and
${\cal C}_2$ as on the other sheets $\Delta_{\theta{\cal C}_i}X=2\pi
\theta v_i$ is automatically satified as $\theta v_i$ belong to the
same coset as $v_i$. 
In order to simplify the integrals, we shall assume that $x$ is real
positive. The general case is easily dealt with.
 The coefficient $a$ and $b$ are determined using 
the periods
$$\eqalign{
\int_{{\cal C}_1}\omega_1&=-2\pi i F(1-x)\cr
\int_{{\cal C}_1}\bar\omega_2&=2\pi i F(1-x)\cr
\int_{{\cal C}_2}\omega_1&=-2\pi i\theta^{1\over 2} F(x)\cr
\int_{{\cal C}_2}\bar\omega_2&=-2\pi i \theta^{1\over 2}F(x)\cr}
\eqno(42)$$
Writing $a=a_1v_1+a_2v_2$ and $b=b_1v_1+b_2v_2$ one gets
$$\eqalign{
a_1&=-b_1={{i}\over 2F(1-x)}\cr
a_2&=b_2={{i\theta^{-{1\over 2}}}\over 2F(x)}\cr}
\eqno(43)$$
The instanton is obtained by integrating (41). Let us first integrate
(41) starting from a real point on the right of $1$. 
One draws an infinitesimal half-circle around each branch point.  
Let us analyse the images of the branch points $1,x,0$ and $\infty$.
 In general  the
instanton is given by 
$$X(z)=2\pi f_2+a_1v_1(\int_{1}^{z}\omega_1-\int_{1}^{z}\bar\omega_2)+
a_2v_2(\int_{1}^{z}\omega_1+\int_{1}^{z}\bar\omega_2)\eqno(44)$$
Now an easy calculation gives 
$$X(x)=2\pi f_2-{{2\pi i}\over \sqrt 3}\theta^{1\over 2}v_1\eqno(45)$$
This point is identified with the fixed point $2\pi f_1$.
Similarly, the image of $0$ is
$$X(0)=2\pi f_1+{{2\pi i\theta^{-{1\over 2}}\over \sqrt 3}}v_2\eqno(46)$$
This is the same fixed point as one has added a vector of the lattice
$L$.
Finally the image of $\infty$ is simply
$$X(\infty)=2\pi f_1+{{2\pi i}\over \sqrt 3}\theta^{1\over
2}v_1\eqno(47)$$
which is nothing but $2\pi f_2$. These four points are the four
vertices of a parallelogram.
 The image of the instanton on the orbifold is then easily
deduced. Notice first that the instanton is a multivalued function on the complex plane. As
one goes around ${\cal C}_{1,2}$, the image of the same point on the
sphere is shifted by $2\pi v_{1,2}$. One can  focus on 
the image of the sphere modulo the lattice generated by $2\pi v_1$ and
$2\pi v_2$. The image is  in a bounded domain. Moreover the
image is repeated three times as one goes from one sheet to
another. Each sheet is sent  to one of the images of the sphere. The images
of the other two sheets is obtained by rotating the image of one sheet by $\theta$ and
$\theta^2$.
Hence we see that the image of the instanton is obtained by
restricting () to one sheet modulo the period lattice generated by
$2\pi v_{1,2}$.
 Once the images of the four marked points is known,
the image of the instanton is the unfolding of the
orbifold passing through the marked points.
 Indeed the image of the instanton is a closed surface,
i.e. a multiple covering of the orbifold. The degree of the instanton
is given by the number of copies of the orbifold necessary to join
the four marked points. One also has to make sure that the complete
image of the instanton obtained by translating the image of one sheet 
by $2\pi(nv_1+mv_2)$ 
form a connected set.   
Several examples are given in Fig.3.
\vskip 1 cm
\midinsert
\epsfysize=5.cm
\epsfxsize=8 cm
{\centerline{\epsfbox{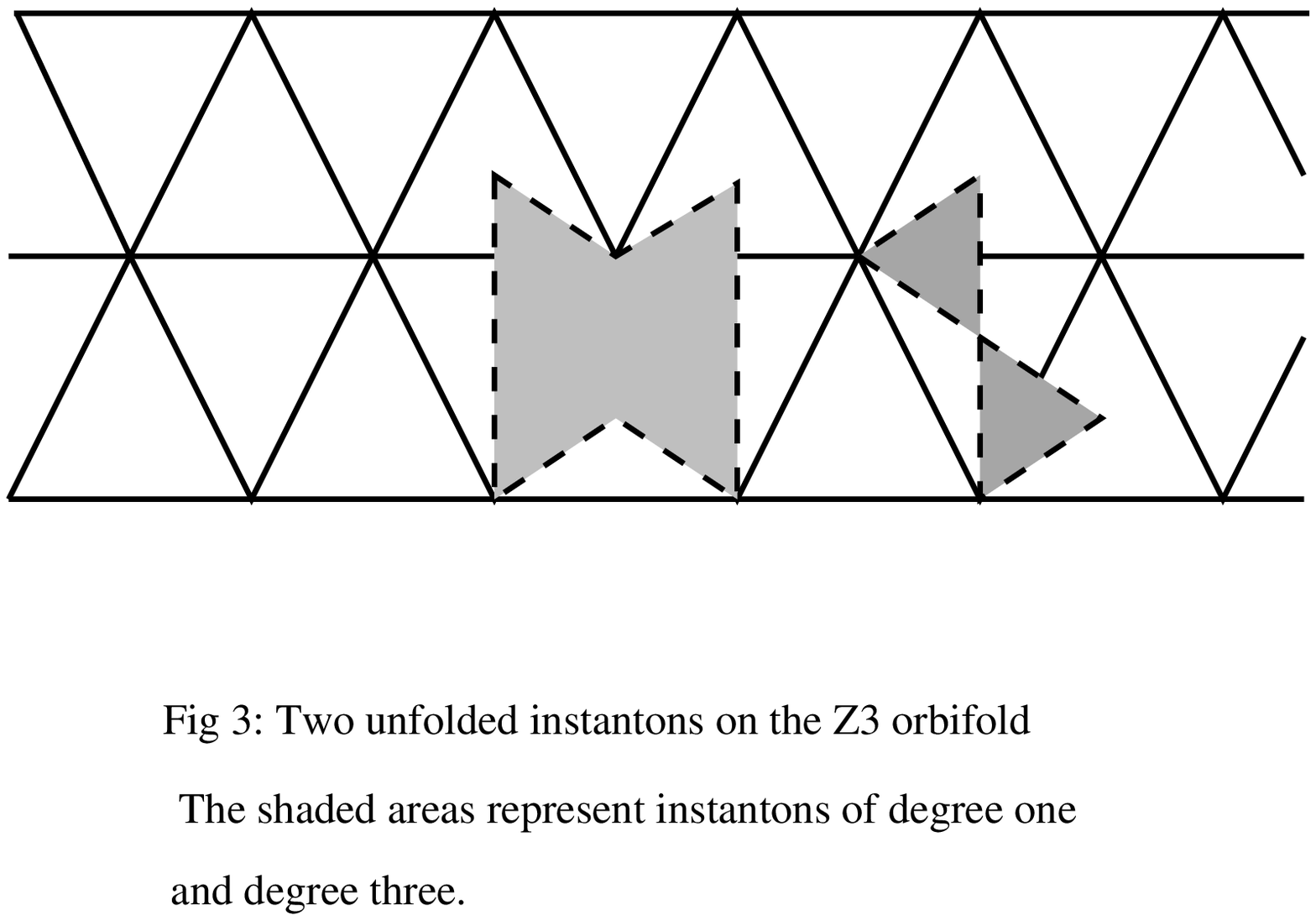}}}
\endinsert
\vskip .5 cm 

We can now calculate the classical action 
using $\int d^2z\vert \omega_{1,2}\vert^2={2{\pi^2}\over \sin
{\pi\over 3}}F(x)F(1-x)$. This yields
$$S={\pi\over 4\hbox{Im}\tau \sin{\pi\over 3}}(\vert v_2\vert^2 + \vert
\tau \vert^2 \vert v_1 \vert^2)\eqno(48)$$
When $x$ is complex one gets
$$S={\pi\over 4\hbox{Im}\tau \sin{\pi\over 3}}(\vert v_2\vert^2
+\hbox{Re}\tau (v_1\bar v_2 \bar \beta +v_2 \bar v_1 \beta )+ \vert
\tau \vert^2 \vert v_1 \vert^2)\eqno(49)$$
where $\beta=-\theta^{_{1\over 2}}$.
The classical part of the correlation function is obtained by summing
over all the boundary conditions specified by $v_1$ and $v_2$.
$$Z_{cl}=\sum_{v_1,v_2}e^{
-{\pi\over 4 Im\tau \sin{\pi\over 3}}(\vert v_2\vert^2
+Re\tau (v_1\bar v_2 \bar \beta +v_2 \bar v_1 \beta )+ \vert
\tau \vert^2 \vert v_1 \vert^2)}\eqno(50)$$ 
The scattering amplitude is given by 
$$T=\int d^2 x Z(x)\eqno(51)$$
We devote the next section to the computation of the scattering
amplitude in the $s$-channel when $x\to 0$. One expects this channel to be sensitive to the exchange of a gauge boson.  We shall see that the $s$ channel bears a perfect analogy with the field theoretic calculations of the previous section.
It is interesting to reformulate the scattering amplitude  in terms of
operators and states in a Hilbert space. This will allow us to
determine the intermediate states dominating  the  scattering
amplitude.
In the limit $x\to 0$ using 
$F(x)\sim 1$ and $F(1-x) \sim -{{\sin {\pi\over 3}}\over \pi}\ln x$
 and  the Poisson resummation formula, one can obtain $Z(x)$ as a sum
over momenta in the dual lattice $p\in L^*$ and winding vectors $v_2\in L$  
$$Z(x)\sim -u\vert x\vert^{-2-s}
\sum_{v_2\in (1-\theta)L,\ p\in L^{*}}
e^{2\pi i(f_1-f_2).p}x^{{{(p+{v_2\over
2})^2}\over 2}}\bar x^{{(p-{v_2\over
2})^2}\over 2}\eqno(52)$$
One can compare this expansion with the result obtained using the
operator product expansion of $\bar {27}(x,\bar x)27 (0)$ and identify
$$\eqalign{
\bar {27}(x,\bar x)27(0)\vert 0>=\sum_{p\in
L^*,\ v_2\in (1-\theta)L} e^{2\pi i(f_1-f_2).p}x^{H_R-{s\over 2}-1}\bar x^{\bar
H_L-{s\over 2}-1}\cr
ik_2.\psi(0)e^{-\phi(0)}e^{i(P_1-P_2).F(0)}
e^{i(k_1+k_2).X(0)} \phi_{v_2,p}(0)\vert 0>\cr}
\eqno(53)$$
The operators $\phi_{v_2,p}(0)$ create states $\vert v_2,p>$ satisfying
$$\eqalign{
H_L\vert v_2,p>={1\over 2}(p-{v_2\over 2})^2\vert v_2,p>\cr
H_R\vert v_2,p>={1\over 2}(p+{v_2\over 2})^2\vert v_2,p>\cr
}\eqno(54)$$
when $H_{L,R}$ are the left and right bosonic Hamiltonians on the orbifold.
 It is  clear that the operators $\phi_{v_2,p}(z,\bar
z)$ are equal to 
$$\eqalign{
\phi_{v_2,p}(z)=e^{i(p+{v_2\over 2}).X(z)}\cr
\phi_{v_2,p}(\bar z)=e^{i(p-{v_2\over 2}).\bar X(\bar z)}\cr}
\eqno(55)$$
where $X_i(z)$ is the six-dimensional field on the orbifold. 
The 
operators $\phi_{v_2,p}$  create winding states on the six dimensional
torus
with momentum $p$ and winding number $2\pi v_2$.
Indeed, let us define the momentum operator as
$$P_i={1\over 4\pi}\int (\partial X_i dz-\bar \partial \bar X_id\bar
z)\eqno(56)$$  
This operator gives the momentum of   strings propagating on
the orbifold. Similarly, the winding number operator is
$$W_i=\int(\partial X_i dz+\bar \partial \bar X d\bar z)\eqno(57)$$
 It is easy to check that
$$\eqalign{
P\vert v_2,p>&=p\vert v_2,p>\cr
W\vert v_2,p>&=2\pi v_2\vert v_2,p>\cr}
\eqno(58)$$
Untwisted solitonic states are created in the
$s$ channel.
\vskip 1 cm
\midinsert
\epsfxsize 8 cm
\epsfysize=7.cm
{\centerline{\epsfbox{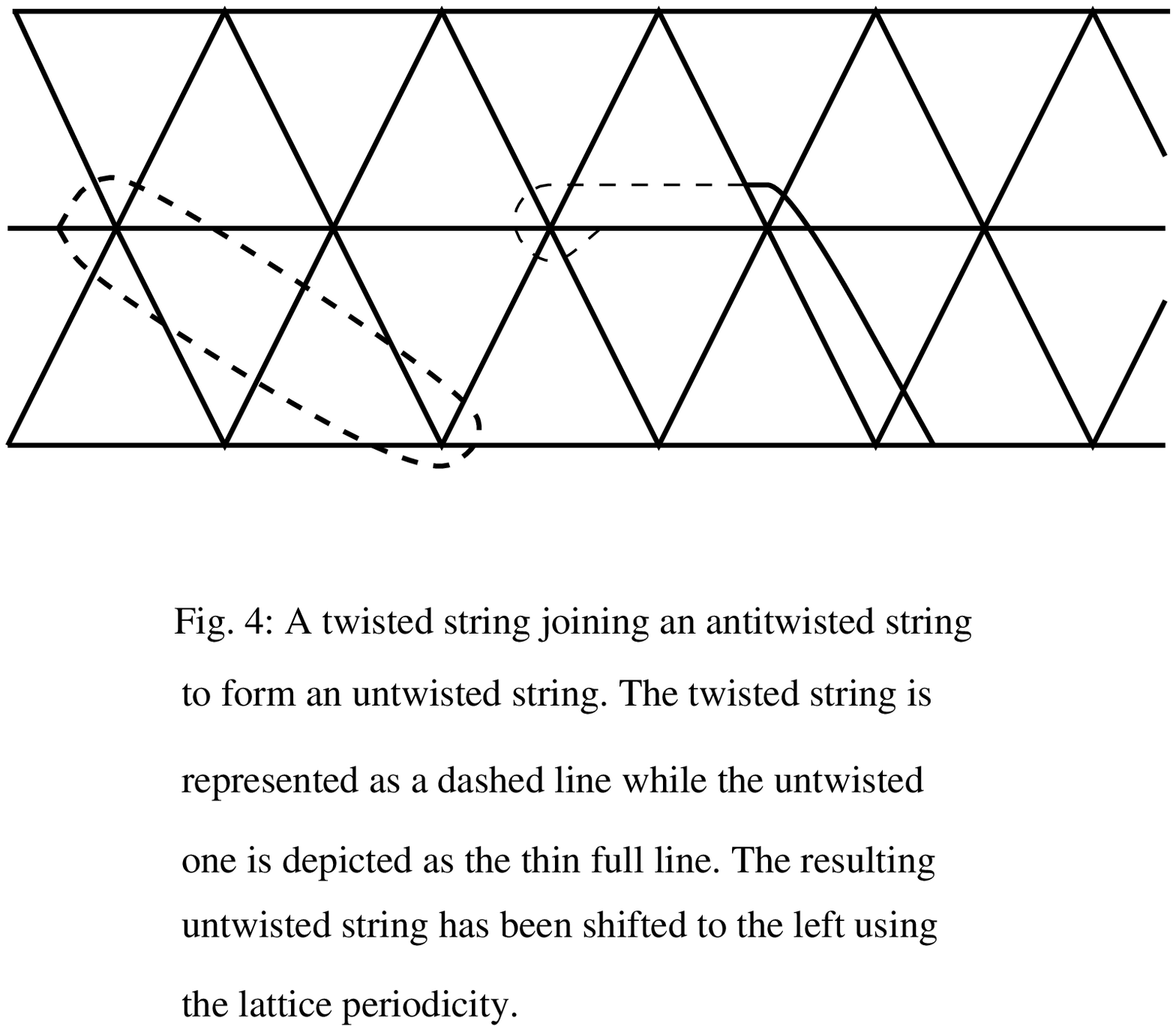}}}
\endinsert
\vskip .5 cm

  The scattering amplitude is
obtained by integrating $Z(x)$ over $x$. Writing $x=\vert
x\vert  e^{i\sigma}$ one gets the integral
$$T\sim -u\int_{x\sim 0}  
d\vert x\vert d\sigma  \sum_{p\in L^*,\
v_2\in (1-\theta)L}e^{2\pi ip.(f_1-f_2)}e^{i\sigma p.v_2}
\vert x\vert ^{p^2+{{\vert v_2\vert^2}\over 4}-1 -s}\eqno(59)$$
where we have used $-u-t=s$.
The $\sigma $ integral gives a Kronecker delta function $\delta
(p.v_2)$.
 Putting $\vert x\vert =e^{-t_E}$, the scattering amplitude becomes
$$T\sim -u\int^{\infty}dt_E \sum_{p\in L^*, v_2\in
(1-\theta)L}\delta (p.v_2)e^{2\pi ip.(f_2-f_1)}e^{-t_E(p^2+{{\vert v_2
\vert^2}\over 4}-s)}\eqno(60)$$
This formulation of the scattering amplitude is readily interpreted if
one introduces the family of Hilbert spaces ${\cal H}_{v_2}$ spanned
by the states $\vert v_2, p,\psi_{\mu}>$ appearing on the right hand side of the operator product expansion (53) for each $v_2$.
It is easily to see that 
$$(L_0+\bar L_0-2)\vert v_2, p,\psi_{\mu}>=(p^2+{{\vert
v_2\vert^2}\over 4}-s)\vert v_2, p,\psi_{\mu}>\eqno(61)$$
as well as 
$$(L_0-\bar L_0)\vert v_2, p,\psi_{\mu}>=(p.v_2)\vert
v_2, p,\psi_{\mu}>\eqno(62)$$
The scattering amplitudes therefore reads
$$T\sim -u\int^{\infty}dt_E  \sum_{v_2\in (1-\theta)L}Tr_{{\cal
H}_{v_2}}( \Gamma_{L_0-\bar L_0}e^{2\pi iP.(f_1-f_2)}e^{-t_E(L_0+\bar
L_0-2)})\eqno(63)$$ 
where $\Gamma_{L_0-\bar L_0}$ is the projector on states satisfying
$L_0=\bar L_0$. We immediately recognise the propagator of strings in
the Hilbert ${\cal H}_{v_2}$ with momentum operator $P$ and
Hamiltonian $L_0+\bar L_0-2$. 
The propagation is in Euclidean time.
Moreover the propagation corresponds to the motion of 10-dimensional
string  from $f_1$ to $f_2$.
One can match this Hamiltonian formulation of the scattering amplitude
with the path integral approach. As for point particles, the
propagation of a  wave packet of states  $\vert v_2,
p,\psi_{\mu}>$ between the fixed points in Euclidean  time is
equivalent to the emission of a classical instanton characterised by
its winding $2\pi v_2$ and surrounded by Gaussian  fluctuations.

The scattering amplitude is similar to the one obtained in the field theoretical case.
 The only difference is the constraint $p.v_2=0$
corresponding to the translation operator along a closed string $U=e^{ i\sigma (L_0-\bar L_0)}$. 
We can now directly apply our analysis of a free particle
compactified on a torus.
Let us rewrite the scattering amplitude by distinguishing the role of
$p=0$ 
$$\eqalign{
&T\sim \sum_{v_2\in (1-\theta)L}{u\over {s-{\vert v_2\vert^2}\over
4}}\cr
&-u\int^{\infty}dt_E \sum_{p\in L^*-\{0\}, v_2\in
(1-\theta)L}\delta (p.v_2)e^{2\pi ip.(f_2-f-1)}e^{-t_E(p^2+{{\vert v_2
\vert^2}\over 4}-s)}\cr}
\eqno(64)$$
As for free particles, one can see that in the vicinity of ${{\vert
v_2\vert^2}\over 4}$ the scattering amplitude is dominated by poles
corresponding to the emission of  winding states of masses ${\vert
v_2\vert^2\over 4}$.
The first pole at $s=0$ is due to the exchange of a
gauge boson of the gauge group $G$ defined by $p=0,\ v_2=0$$^{[9]}$. 
The other poles are all due to winding massive states. The sign of the pole is consistent with the result $T\sim (-1)^{S+1}{{u}\over s-m^2}$ for the exchange of a particle of spin $S$ and mass $m$. Here one knows that $S=1$ for vector bosons.
Notice that these poles are not suppressed by the volume of the orbifold.
Indeed, the normalisation is such that the effect of  the gauge boson
is still present is the large radius limit.

 Away from these poles, we know that for each
winding vector $v_2$ the behaviour of the partial scattering amplitude
$T(v_2)$ depends on the energy $s$. Below the pole, the scattering
amplitude is dominated by a saddle point corresponding to an instanton
of winding vector $v_2$ while above the pole the saddle point point is
due to a soliton. The only difference with the free particle case comes
from the GSO projection. Using the poisson resummation formula, we get
$$T\sim -u V_L \int^{\infty}dt_E\int_0^{2\pi}d\sigma {1\over
t_E^3}\sum_{q_1\in L, v_2\in (1-\theta)L'}e^{-t_E({{\vert
v_2\vert^2\over 4}-s})}e^{-{{2\pi (f_1-f_2-q_1+\sigma v_2)}^2\over
4t_E}}\eqno(65)$$
where we have subtracted the contribution of the gauge boson for $v_2=0$ as $L'=L-\{ 0\}$. 
Let us concentrate on one particular $v_2$. Below the pole, the
corresponding winding mode contributes as
$$T(v_2)\sim -u V_L\int_{0}^{2\pi}d\sigma {1\over ({{\vert v_2\vert^2}\over
4}-s)^3} e^{-2\pi \sqrt {{{\vert v_2\vert^2}\over
4}-s}\vert f_1-f_2-q_1+\sigma v_2\vert}\eqno(66)$$
where we have used a saddle point approximation as explained the appendix.
Notice that the scattering amplitude is exponentially suppressed. This result is strikingly similar to the field theoretical result. The only difference comes from the average over the closed string parameter $\sigma$.
 It is interesting to remark that the saddle  point
equation reads 
$$s+P^2={{\vert v_2\vert^2}\over 4}\eqno(67)$$
where $P$ is the six-momentum of the instanton. The mass of each of these instantons is nothing but the mass of the corresponding quantum state propagating between the fixed points, i.e. ${\vert v_2\vert }\over 2$. 

Let us come back on the resulting coupling in the effective field theory. We are interested in reproducing the above scattering amplitude in an effective theory. This effective theory depends on the scale at which one identifies the string scattering amplitude on the sphere with a coupling in the effective Lagrangian. The string calculation is performed at a scale $s\sim M^2_c$ close to the compactification scale. Dimensionally one has $M_c\sim {M_{pl}\over R}$ where $R$ is the radius of the orbifold in Planck units. Similarly the masses of the Kaluza-Klein modes is proportional to $ \vert v_2\vert \sim RM_{pl}$. One can therefore neglect $s$ compared to $\vert v_2\vert^2 $ for reasonable values of $R\ge 1$. Fixing the value of $s=M_c$ we find that there is a coupling in the effective Lagrangian of the form
$$\Delta V= {T\over M_{pl}^2}(D_{\mu}\bar \phi_1 T^a \phi_1)(\bar \phi_2 T^a D^{\mu}\phi_2)\eqno(68)$$
where $D$ is the covariant derivative.
The coupling contant $T$ behaves like 
$$T\sim V_L\sum_{v_2\in (1-\theta)L-\{ 0\}}
\int_{0}^{2\pi}d\sigma{1\over \vert v_2\vert^6}
 e^{-\pi  {{{\vert v_2\vert}}}\vert f_1-f_2-q_1+\sigma v_2\vert}\eqno(69)$$
 This new coupling is exponentially suppressed. The argument of the exponential is simply the mass of the Kaluza-Klein modes multiplied by the distance between the fixed points. This distance depends on the number of times the solitonic states wrap around the orbifold. Dimensionally, the exponent varies as $R^2$ where $R$ is the radius of the orbifold. We have  obtained contributions to the  potential which are similar to the ones of the field theoretic compactification on a singular torus. Due to supersymmetry this exponentially small term modifies the Kahler potential, i. e. the kinetic terms. Notice that the coupling appears with two derivatives and is therefore suppressed by the Plank mass. The main difference between the field theoretic and stringy calculations stems from the nature of the intermediate states. In the field theoretic example, the exchanged particle is a Lorentz scalar whereas in the stringy case it is a Lorentz vector.
This entails that the string coupling is derivative. It would be conspicuous to know if higher order scattering amplitudes yield non-derivative couplings occurring in the scalar potential.
\vskip 1 cm
\vskip 1 cm
\leftline{\bf VI Conclusion}
\vskip 1 cm
We have been interested in naturally small couplings between massless states in effective field theories. We have shown that exponentially small couplings appear when massless fields located at singular points of a compactified space are coupled to massive states. These massive states propagate between the singular points leading to couplings  depending exponentially on the masses of the intermediate states and the distance between the singularities. An interesting framework for this type of mechanism is provided by the heterotic string theory compactified on an orbifold. In that case the twisted states are located at the fixed points of the orbifold. These fixed points are singular. There exists an infinite tower of untwisted massive solitonic states coupling the fixed points. These Kaluza-Klein modes give rise to exponentially small couplings in the effective field theory.
The existence of exponentially small couplings may be relevant to the hierarchy problem. Indeed one can hope that the large difference between the Planck scale and the weak scale is due to some small coupling whose origin would be a sign of  extra stringy dimensions.  

\vskip 1 cm
\leftline{\bf Appendix}
\vskip 1 cm

In this appendix we shall compute the scattering amplitude $\phi_1+\phi_1 \to \phi_2+\phi_2$  in the case $D=1$. In particular, we give a path integral formulation which reappears in the stringy calculation.
One needs to evaluate the Green function
of the operator $\partial_A^2+m^2$. Suppose that the  $d$
dimensional momentum of the particle is fixed  $s=p^2$. One is
therefore left with the operator $m^2-s-\partial_{d+1}^2$.
The propagator reads
$$<f_2\vert ({{m^2-s}\over 2}-{{\partial_{d+1}^2}\over 2})^{-1}\vert
f_1>=<f_2\vert \int_0^{\infty}e^{-{t\over 2} 
(m^2-s-\partial_{d+1}^2)}\vert f_1>\eqno(A1)$$ 
The evolution operator $e^{t{{\partial_{d+1}^2}\over 2}}$ gives the
solution of the heat equation ${{\partial }\over \partial t}\vert
\psi>=-H\vert \psi >$ where $H=-{{\partial_{d+1}^2}\over 2}$. 
This can be written
as the path integral in euclidean time $t$
$$<f_2\vert ({{m^2-s-\partial_{d+1}^2}\over 2})^{-1}\vert
f_1>=\int_0^{\infty}e^{-t{{(m^2-s)}\over 2}} \int_{f_1}^{f_2}{\cal
D}x(t)e^{-{1\over 2}\int_0^t\dot x^2 d\tau}\eqno(A2)$$
The path integral corresponds to Brownian trajectories from
$f_1$ to $f_2$.  
 To evaluate this  path integral, we shall use a saddle point
approximation.
First of all, the path integral is dominated by paths satisfying the
equation of motion
$$\ddot x=0\eqno(A3)$$
This is the equation of motion of a point particle in euclidean time,
i.e. an instanton going from $f_1$ to $f_2$.
 This instanton is
therefore
$$x(\tau)=f_1+p\tau\eqno(A4)$$
where
$$p={{f_2-f_1}\over t} \eqno(A5)$$
Substituting in the Green function, one gets
$$<f_2\vert (m^2-s-\partial_{d+1}^2)^{-1}\vert
f_1>={1\over 2}\int_{0}^{\infty} dt e^{-t{{(m^2-s)}\over 2}-{1\over 2}{{\vert
f_2-f_1\vert ^2}\over t}}\int_0^0 {\cal D}x(t)e^{-{1\over 2}\int_0^t
d\tau \dot x^2 d\tau}\eqno(A6)$$
The path integral concerns closed trajectories returning to the origin
in a time $t$, its value is $(\sqrt {2\pi t})^{-1}$.
The $t$ integral can be evaluated using a saddle point approximation.
First of all, notice that this integral
converges when $s<m^2$, i.e.  the energy of the particle is not large
enough to create a real particle. In that case the propagator between
$f_1$ and $f_2$ corresponds to the emission of an instanton. The
saddle point equation reads
$$m^2-s -{{\vert f_2-f_2\vert^2}\over t^2}=0\eqno(A7)$$
which implies that 
$$t={R\over \sqrt {m^2 -s}}\eqno(A8)$$
Using this result, it is easy to rewrite the saddle point equation
$$s+p^2=m^2\eqno(A9)$$
This is the mass relation for an instanton of mass $m^2$ and Euclidian
$(d+1)$ momentum $p$. We can now get the propagator
$$<f_2\vert (m^2-s-\partial_{d+1}^2)^{-1}\vert
f_1>\sim {1\over \sqrt {m^2-s}}e^{-\sqrt {m^2 -s}\vert f_2
-f_1\vert} \eqno(A10)$$
We see that the saddle point approximation of the Green function gives
for the scattering amplitude  
$T=-\lambda^2 <f_2\vert (m^2-s-\partial_{d+1}^2)^{-1}\vert
f_1>$.
Moreover, one can interpret this decay as the
propagation of an instanton between the two fixed points.

Let us now analyse the opposite situation when $s>m^2$. In that case
the previous integrals are not convergent. One can nevertheless
analytically continue the propagator  which is now dominated by the
exchange of a real particle, i.e. a soliton. 
The contour of integration can be deformed to the imaginary axis
$$<f_2\vert ({{m^2-s-\partial_{d+1}^2}\over 2})^{-1}\vert
f_1>=\int_{0}^{i\infty}dt
e^{-t{{m^2-s-\partial_{d+1}^2}\over 2}}\eqno(A11)$$
when  $s<m^2$ as $-\partial_{d+1}^2$ is a  positive operator.
Let us now put $t=i\tau$ where $\tau$ is a real time parameter. This
implies that
$$<f_2\vert ({{m^2-s-\partial_{d+1}^2}\over 2})^{-1}\vert
f_1>=i\int_{0}^{\infty} d\tau e^{-i\tau{{(m^2-s)}\over
2}} e^{i\tau{{\partial_{d+1}^2}\over 2}}\eqno(A12)$$
The evolution operator $e^{i\tau{{\partial_{d+1}^2}\over 2}}$ corresponds to
the Schrodinger equation $i{{\partial}\over \partial \tau}\vert \psi
>=H\vert \psi >$ where $H=-{{\partial_{d+1}^2}\over 2}$. One can
 write the evolution operator as a path integral 
$$<f_2\vert ({{m^2-s-\partial_{d+1}^2}\over 2})^{-1}\vert
f_1>=i\int_{0}^{\infty} d\tau e^{-i\tau{{(m^2-s)}\over
2}} \int_{f_1}^{f_2}{\cal D}x(\tau)e^{{i\over 2}\int_0^{\tau}\dot x^2
d\tau_0} \eqno(A13)$$
This is the representation of the propagator when $s>m^2$
Notice that now the sum is over classical trajectories of a real
particle. As before the path integral is dominated by solitons
satisfying
$$\ddot x=0\eqno(A14)$$
whose solution is
$$x=f_1+p\tau_0\eqno(A15)$$
where $p={{f_2-f_1}\over \tau}$. The propagator becomes
$$<f_2\vert ({{m^2-s-\partial_{d+1}^2}\over 2})^{-1}\vert
f_1>=i\int_{0}^{\infty} d\tau (2\pi i \tau )^{-{1\over 2}} 
e^{-i\tau{{(m^2-s)}\over
2}+{i\over 2}p^2\tau}\eqno(A16)$$
The saddle point equation becomes 
$$m^2-s +{{\vert f_2-f_1\vert}^2\over \tau^2}=0\eqno(A17)$$
whose solution is given by 
$$\tau={{\vert f_2-f_1\vert}\over \sqrt {s-m^2}}\eqno(A18)$$
Notice that the saddle point equation reads
$$s-p^2=m^2\eqno(A19)$$
the mass relation for a particle of mass $m$. Hence the propagator
corresponds to the emission of a real particle between $f_1$ and
$f_2$. The integral is obtained after deforming the integration
contour. It goes through the saddle point, the angle between the real
line and the contour being $-{\pi\over 4}$. 
The propagator becomes
$$<f_2\vert ({{m^2-s-\partial_{d+1}^2}\over 2})^{-1}\vert
f_1>\sim (s-m^2)^{-{1\over 2}}e^{i\sqrt {s-m^2}\vert f_2
-f_1\vert} \eqno(A20)$$
The same results appear in the context of orbifolds.

\vfill\eject
\centerline{\bf References}
\vskip 1 cm
\leftline{[1] "Naturalness, chiral symmetry and spontaneouys chiral symmetry breaking}
\vskip .2 cm
\leftline{by G. 't Hooft in $recent\ developments\ in\ gauge\ theories $ carg\`ese  1979 }
\vskip .2 cm
\leftline{eds G. 't Hooft et al. N-Y Plenum.}
\vskip .2 cm
\leftline{[2] N. Turok Phys. Rev. Lett. {\bf 76} (1996) 1015.} 
\vskip .2 cm
\leftline{[3] L. Dixon, J. Harvey, C. Vafa and E. Witten Nucl. Phys {\bf B261} (1985) 651, {\bf B274} (1986) 285.}
\vskip .2 cm
\leftline{[4] S. Hamidi and C. Vafa Nucl. Phys. {\bf B279} (1987) 465.}
\vskip .2 cm
\leftline{[5] L. Dixon, D. Friedan, E. Martinec  and S. Shenker Nucl. Phys. {\bf B282} (1987) 13.}
\vskip .2 cm
\leftline{[6] A. Font, L. E. Ibanez, H. P. Nilles and F. Quevedo Phys. Lett. {B213} (1988) 274.}
\vskip .2 cm
\leftline{[7] A. Font, L. E. Ibanez , F. Quevedo and A. Sierra Nucl. Phys {\bf B331} (1990) 421.}
\vskip .2 cm
\leftline{[8] L. E. Ibanez, H. P. Nilles and F. Quevedo Phys. Lett. {\bf B187} (1987) 25.}
\vskip .2 cm
\leftline{[9] L. Dixon, V. Kaplunovski and J. Louis Nucl. Phys. {\bf B329} (1990) 27.}
\end